\documentclass[conference]{IEEEtran}
\IEEEoverridecommandlockouts
\usepackage{graphicx}
\usepackage{subcaption}
\usepackage{algorithm}
\usepackage{algpseudocode}
\usepackage{amsmath}
\usepackage{textgreek}

\newcommand\nextToken\relax
\usepackage{multirow}
\usepackage{gensymb}
\usepackage{tikz}
\usetikzlibrary{calc}
\usepackage{tabularx}

\newcommand\copyrighttext{%
  \footnotesize \textcopyright 2024 IEEE. Personal use of this material is permitted.
  Permission from IEEE must be obtained for all other uses, in any current or future
  media, including reprinting/republishing this material for advertising or promotional
  purposes, creating new collective works, for resale or redistribution to servers or
  lists, or reuse of any copyrighted component of this work in other works.}
\newcommand\copyrightnotice{%
\begin{tikzpicture}[remember picture,overlay]
\node[anchor=south,yshift=10pt] at (current page.south) {\fbox{\parbox{\dimexpr\textwidth-\fboxsep-\fboxrule\relax}{\copyrighttext}}};
\end{tikzpicture}%
}

\def\BibTeX{{\rm B\kern-.05em{\sc i\kern-.025em b}\kern-.08em
    T\kern-.1667em\lower.7ex\hbox{E}\kern-.125emX}}
\begin{document}

\title{A Data-Driven Analysis of Vulnerable Road User Safety in Interaction with Connected Automated Vehicles}
\author{\IEEEauthorblockN{Edmir Xhoxhi}
\IEEEauthorblockA{
\textit{Institute for Communications Technology}\\
\textit{Leibniz University Hannover}\\
Hannover, Germany \\
edmir.xhoxhi@ikt.uni-hannover.de}\\
\and
\IEEEauthorblockN{Vincent Albert Wolff}
\IEEEauthorblockA{
\textit{Institute for Communications Technology}\\
\textit{Leibniz University Hannover}\\
Hannover, Germany \\
vincent.wolff@ikt.uni-hannover.de}\\
}

\maketitle
\copyrightnotice
% Context: Provide background information for less specialized readers, establishing the importance of the problem.
% Need: State the difference between the desired and actual situations to motivate the audience.
% Task: Describe what the authors undertook to address the need, using the first person (we) and past tense.
% Object: Clarify what the paper covers without repeating the task, using the active voice and present tense.
% Findings: State the main results in a way that is helpful for both less and more specialized readers.
% Conclusion: Interpret the findings and state the implications or recommendations.
% Perspectives: Broaden the view with any further needs and tasks.
%\vspace{-0.5cm}
\begin{abstract}
According to the World Health Organization, the involvement of Vulnerable Road Users (VRUs) in traffic accidents remains a significant concern, with VRUs accounting for over half of traffic fatalities.
The increase of automation and connectivity levels of vehicles has still an uncertain impact on VRU safety.
By deploying the Collective Perception Service (CPS), vehicles can include information about VRUs in Vehicle-to-Everything (V2X) messages, thus raising the general perception of the environment.
Although an increased awareness is considered positive, one could argue that the awareness ratio, the metric used to measure perception, is only implicitly connected to the VRUs' safety.
This paper introduces a tailored metric, the Risk Factor (RF), to measure the risk level for the interactions between Connected Automated Vehicles (CAVs) and VRUs.
By evaluating the RF, we assess the impact of V2X communication on VRU risk mitigation.
Our results show that high V2X penetration rates can reduce mean risk, quantified by our proposed metric, by up to 44\%.
Although the median risk value shows a significant decrease, suggesting a reduction in overall risk, the distribution of risk values reveals that CPS's mitigation effectiveness is overestimated, which is indicated by the divergence between RF and awareness ratio.
Additionally, by analyzing a real-world traffic dataset, we pinpoint high-risk locations within a scenario, identifying areas near intersections and behind parked cars as especially dangerous.
Our methodology can be ported and applied to other scenarios in order to identify high-risk areas.
We value the proposed RF as an insightful metric for quantifying VRU safety in a highly automated and connected environment.
\end{abstract}


\section{INTRODUCTION}

The Day 2 Vehicle-to-Everything (V2X) services have now reached the closing phase of the standardization process under the European Telecommunications Standards Institute (ETSI).
An important aspect of these services is the safety of Vulnerable Road Users (VRUs).
One method to enhance the visibility of VRUs is to allow them to directly share data regarding their kinematic status and position.
The VRU Awareness Service (VAS) \cite{etsi:vam} serves this purpose.
VRUs equipped with devices providing connectivity, such as smartphones or on-board communication units of bicycle or motorbike, disseminate VRU Awareness Messages (VAMs) themselves in order to announce their dynamics periodically to their environment, more specifically Intelligent Transportation System Stations (ITS-S) such as connected vehicles or road side units.
The Collective Perception Service (CPS) \cite{etsi:cpm} on the other hand lets connected vehicles share information about other road users between each other, utilizing the information gathered by the onboard sensors of vehicles.
By implementing CPS and VAS on ITS-S, vehicles that are beyond the line of sight can still be aware of traffic participants, especially VRUs such as cyclists, motorcyclists, or pedestrians.
The received information enriches the local environmental model of Connected Automated Vehicles (CAVs) to adjust their planned trajectory accordingly.

The World Health Organization (WHO) warns that car accidents are the leading cause of death among young individuals.
In their most recent report, it is stated that more than half of the fatalities on the road involve VRUs, be they pedestrians, cyclists, or bikers \cite{injuriesWHO}.
These numbers highlight the importance of advancing new technologies that enhance VRU safety.
The authors in \cite{puller2021towards} and \cite{morgenroth2009improving} conduct analytical studies on the reasons for accidents, based on datasets.
Puller et al., the authors of the former study, list random road crossing by pedestrians as one of the main causes of their involvement in traffic accidents.
Morgenroth et al. add occlusion by objects, such as parked cars on the side of the road, as another cause of accidents.
With their sight being occluded, drivers have very little time to adjust and react to VRUs entering their vehicle's trajectory.

In this study, we introduce a metric designed to assess the risk of interactions between VRUs and Automated Vehicles (AVs), specifically focusing on the increased risk due to occlusion. This metric, with necessary modifications, can also be used to evaluate the risk involved in interactions between VRUs and conventional vehicles. A key aspect of this Risk Factor (RF) is its ability to assess the risk at specific locations or points in any scenario by analyzing relevant data, enabling a detailed evaluation of risk across different traffic conditions. This metric not only improves our understanding of VRU safety but also serves as an important tool for traffic management and urban planning.

Furthermore, we apply this metric in a simulation that utilizes real-world data to gather realistic measurements and explore the influence of V2X technology on the safety of VRUs. By employing these datasets, we perform a detailed examination of risk zones at specific intersections, demonstrating the impact of obstructions on VRU safety. This part of the research emphasizes how V2X technology can decrease the dangers posed by restricted visibility and various environmental conditions. The results offer significant insights into the ways V2X communication can improve situational awareness and lessen the chances of incidents involving VRUs, thereby aiding in the creation of safer urban traffic conditions.

% The remainder of this paper is structured as follows: The following section offers an overview of related work, highlighting gaps in the existing literature and emphasizing our contributions.
% Section \ref{sec:riskFactor} details the RF as a metric for quantifying risk in interactions between CAVs and VRUs.
% Section \ref{sec:furtherDiscuss} provides additional explanations and further discussion on the application of the RF.
% Then the architecture and parameters of the simulation framework is introduced.
% The findings and conclusions of this study are respectively presented in the sixth and seventh sections.

\section{RELATED WORK}
\label{sec:relatedWork}
In the following Related work in the context of VRU safety and connectivity considering VRUs is presented.
The authors in \cite{morgenroth2009improving}\cite{puller2021towards} focus on an analytical aspect of VRU risk, primarily concentrating on the causes of accidents.
Meanwhile the authors in \cite{shalev2017formal} postulate Responsibility-Sensitive Safety set of rules which can be used to formalize the safe operation of automated vehicles.
Khayatian et al. \cite{khayatian2021cooperative} extend these set of rules in order to accommodate CAVs.

Since the safety of VRUs is an important topic and we are experiencing an increase in automation levels, it is important to define metrics and parameters which quantify the interactions of these category of traffic participants.
Schiegg et al. examine the safety parameters that can be utilized for collective perception during the operation of CAVs \cite{schiegg2021collective}.
Collective perception serves as an effective bridging point between CAVs and VRUs. The former can use their onboard sensors to detect the latter and then transmit this information via Collective Perception Messages (CPMs).
In the study mentioned above, the metrics measured on the higher layers are particularly interesting because they have a direct impact on the user's experience. The metrics mentioned in the study are Environmental Risk Awareness, Time-to-Plan (TTP), and the Comprehensive Safety Metric.
Banjade et al. \cite{Banjade2021Vulnerable} adapt the TTC metric to mitigate false triggers of collision warnings, introducing Minimum Safe Distance (MSD) metric, which considers lateral, longitudonal and vertical distance. By implementing VAS, they show the robustness of the proposed metric.
Karoui et al. \cite{karoui2022systems} adopt a more practical approach in considering the effect of V2X communication on the safety of VRUs.
In their study, they examine a single scenario where a VRU emerges from an occluded area, taking into account the probability of the driver's awareness depending on the technology used or the message sent.
Working in this direction, and adopting a more holistic view, the authors in \cite{lobo2022enhancing} and \cite{willecke2021vulnerable} explore the impact of V2X communication on VRU safety:
Lobo et al.\cite{lobo2022enhancing} investigate the effect of combining VAM and CPM messages.
To compare and evaluate the results, the authors utilized VRU perception rate and detection time.
Combining VAM and CPM appears to be the best solution, improving both the perception rate and detection time while maintaining the channel load in a relaxed state.
Willecke et al. \cite{willecke2021vulnerable}, on the other hand, examine the effect of including VRUs in CPM messages.
As metrics for evaluating different settings, they use the awareness ratio within a 25 m radius and the detection delay. The latter refers to the time it takes for a vehicle to become aware of a VRU that will be entering the radius.

In this literature review, it is noticeable that metrics such as the awareness ratio or versions of the TTP metric are predominantly used.
In this work, we propose a metric specifically designed for assessing the interactions between CVs and VRUs. While the awareness ratio is useful for highlighting how occlusion can be mitigated through the sharing of sensor data via connectivity, it falls short in adequately considering the planned paths of the traffic participants.
For instance, a VRU occluded behind a vehicle would be marked as occluded and thus perceived as a risk by the awareness ratio algorithm. However, this assessment does not align with reality if the vehicle is moving away from the VRU. To address this limitation, we introduce the RF metric, which will be elaborated upon in the following section. This metric takes into account the planned trajectories, thereby providing a more accurate quantification of risk.
\begin{figure}[!b]
  \centering
  % First subfigure
  \begin{subfigure}[b]{\linewidth}
    \includegraphics[width=\linewidth]{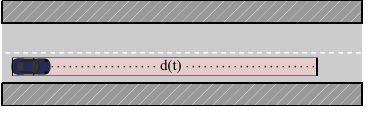}
    \caption{Automated vehicle with planned trajectory}
    \label{fig:ear_low}
  \end{subfigure}
  \begin{subfigure}[b]{\linewidth}
    \includegraphics[width=\linewidth]{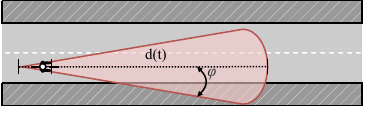}
    \caption{Two-wheeled VRU}
    \label{fig:ear_high}
  \end{subfigure}
  \label{fig:ear}
    \begin{subfigure}[b]{\linewidth}
    \includegraphics[width=\linewidth]{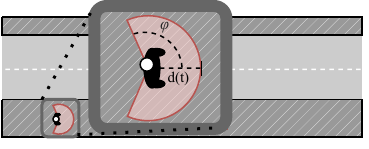}
    \caption{Pedestrian}
    \label{fig:ear_high}
  \end{subfigure}
  \caption{Risk area for different types of traffic participants}
  \label{fig:riskArea}
\end{figure}
\section{Risk Factor}
\label{sec:riskFactor}
In this section, we describe the Risk Factor (RF) metric utilized in our study.
This metric is specifically designed for assessing the interaction risk between CAVs and VRUs.
However, with certain modifications, it can also be applied to risk assessment involving legacy vehicles and other traffic participants.
The RF metric is conceptualized as an overlap of two key metrics: TTC and TTP.
The former calculates the minimal time until two moving entities collide, based on their current trajectories and speeds.
The TTC metric typically uses a one-track prediction model for both actors, which assumes that each actor will continue moving along their current path at a constant speed. The ETSI defines the TTC particularly for the interaction between one vehicle and VRU \cite{etsi:vru_safety}.
If the trajectories of the two actors intersect, the TTC metric will return the time until their collision.
If their paths do not intersect, the TTC will return infinity, indicating that a collision is not expected based on the current information.
TTP on the other hand is defined as the window of opportunity between the moment the detection of a traffic participant and the moment when an action must be taken to avoid or mitigate a collision.

The proposed RF metric is proposed to assess risk, taking into account the unique interactions between VRUs and automated vehicles. It incorporates the potential for swift changes in the trajectories of VRUs by establishing a zone that predicts where the VRU will most likely move within the next 5 seconds, factoring in potential changes in its heading angle.
The innovative aspect of the RF metric lies in its ability to be mapped to a single comprehensive normalized value.
This characteristic simplifies the process of comparison and analysis.
Another significant feature of the RF metric is its capability to be calculated live from the CAVs, similar to TTP.
This real-time calculation allows for dynamic and immediate risk assessment in varying traffic scenarios, enhancing the overall utility and applicability of the metric in real-world situations.

Before continuing with further discussions about the impact of this metric, we will first describe how to calculate the RF and the assumptions made during the process.
The first assumption is that AVs will be able to know their planned trajectory for the next 5 seconds.
Second, if a VRU is perceived from the sensors, its speed and direction can be also considered as known, which is not an unusual requirement for modern environmental on-board sensors.
In Figure \ref{fig:riskArea} we show the risk area for an AV and VRUs, which is an important concept for the explanation of the RF.
The risk area for the AV is similar to the planned trajectory and extends in width to match the vehicle's dimensions.
The length of the risk area and the position of the AV within the risk area is a function of time.
As for the VRUs, the risk area is calculated from the AV perspective.
The opening angle \textphi, which is measured from the perceived direction angle of the VRU and leaves room for uncertainties in the path prediction of VRUs.
The length of the risk area d(t) and the position within the risk area is calculated by the formula $d=\frac{v_{act}}{t}$, where $v_{act}$ is the actual speed.
\begin{figure}[t]
    \centering
    \includegraphics[width=\linewidth, trim=0cm 0cm 0cm 0.4cm, clip]{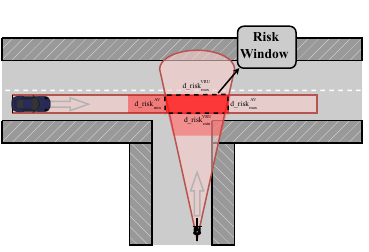}
    \caption{Depiction of risk scenario between an AV and a VRU}
    \label{fig:riskwindow}
\end{figure}
With that in mind, we can define the risk window $W_r= [d_{risk, min}\quad d_{risk, max}]$, as shown in Figure \ref{fig:riskwindow}.
When speaking about safety, distances should be expressed as a function of time since traffic participants are travelling at different speeds. We make the conversion of $d_{min}$ and $d_{max}$ with the help of the following formulas:
\begin{align}
&t_{risk, min} = \frac{d_{risk, min}-length}{v}\\
&t_{risk, max} = \frac{d_{risk, max}}{v}
\end{align}
Note that for $t_{risk, min}$, we include the length of the traffic participants, whether it be an AV or a VRU. This approach is adopted because the measurement of distance originates from the rear end of the traffic participants, and a collision between the AV and VRUs is possible whenever any part of the VRU enters the risk zone.
The risk window in the time domain will be given as: \(T_{r} = [ t_{\text{risk, min}}, t_{\text{risk, max}} ]\).

Returning to the example in Figure \ref{fig:riskwindow}, the AV, shown in dark blue, will calculate the risk window for itself \(T_{r}^{\text{AV}}\) and the risk window for the cyclist \(T_{r}^{\text{VRU}}\) as soon as it will be able to sense it.

%The risk window in the time domain will be given as: $T_{r} = [ t\_risk_{min}\quad t\_risk_{max} ]$.
%Returning to the example in Figure \ref{fig:riskwindow}, the AV, shown in dark blue, will calculate the risk window for itself $\sideset{T}{_r}{^{AV}}$ and the risk window for the cyclist $\sideset{T}{_r}{^{VRU}}$ as soon as it will be able to sense it.
This sensing is done either via on board sensors, or connectivity.
The risk time (RT) will be than calculated by the formula:
%\begin{align}
%RT = 
%\begin{cases}
%min( \sideset{T}{_r}{^{AV}} \cap \sideset{T}{_r}{^{VRU}} ), & \text{if } \sideset{T}{_r}{^{AV}} \cap \sideset{T}{_r}{^{VRU}} \neq \phi \\
%max\_val & \text{else}
%\end{cases}
%\label{eq:RT}
%\end{align}
\begin{align}
RT = 
\begin{cases}
\min(T_r^{AV} \cap T_r^{VRU}), & \text{if } T_r^{AV} \cap T_r^{VRU} \neq \emptyset \\
\infty & \text{else}
%\text{max\_val} & \text{else}
\end{cases}
\label{eq:RT}
\end{align}

\begin{figure}[t]
    \centering
    \includegraphics[width=\linewidth, trim=0.7cm 0cm 0.7cm 1.2cm, clip]{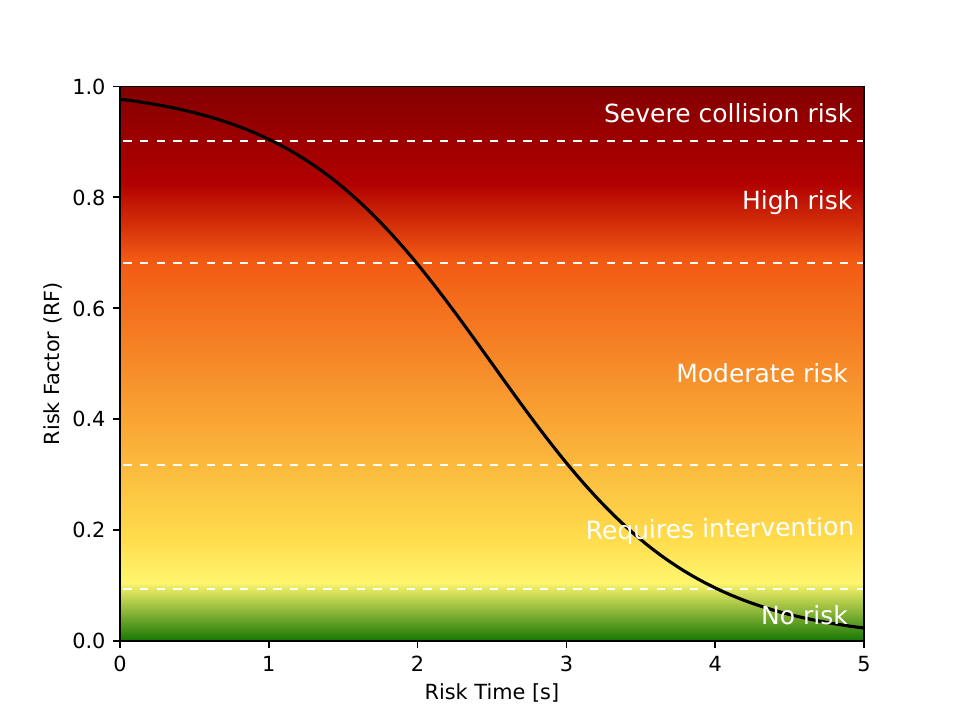}
    \caption{Risk factor as a function of risk time}
    \label{fig:riskfinal}
\end{figure}

%According to the Equation \ref{eq:RT}, risk time is the minimal value of the intersection between $\sideset{T}{_r}{^{AV}}$ and $\sideset{T}{_r}{^{VRU}}$, if the intersection segment is not empty.
According to the Equation \ref{eq:RT}, risk time is the minimal value of the intersection between \(T_r^{AV}\) and \(T_r^{VRU}\), if the intersection segment is not empty.
On the other hand, if the intersection between the windows is empty, the RT is set to an arbitrarily high number.
The higher the RT is, the lower the risk of encounter is.
This calculation of RT is very similar to TTC except for the second condition.
We are able to define RT with an arbitrarily high number because of the mapping function that we use next.
In the last step we define the RF by using the formula:
\begin{align}
    RF = sigmoid(\alpha \times (RT - \tau))
    \label{eq:RF}
\end{align}
We choose $\alpha=-1.5$ and $\tau=2.5 sec$ which builds the function shown in Figure \ref{fig:riskfinal}.
% \begin{table}[htbp]
%     \centering
%     \begin{tabular}{|c|c|}
%         \hline
%         RT Value & RF Value \\
%         \hline
%         0.5 & 0.95 \\
%         \hline
%         1.0 & Column 6 \\
%         \hline
%         1.5 &  0.73 \\
%         \hline
%         2.0 & 0.5 \\
%         \hline
%         2.5 &  0.27 \\
%         \hline
%         3.0 & 0.12 \\
%         \hline
%         4.0 & 0.02\\
%         \hline
%     \end{tabular}
%     \caption{A 2x6 Table}
%     \label{tab:mytable}
% \end{table}
% Values for sigmoid 1 / (1 + np.exp(2.5 * (x - 2)))
%x=0, y=0.9770226300899744
%x=1, y=0.9046505351008906
%x=2, y=0.679178699175393
%x=3, y=0.3208213008246071
%x=4, y=0.09534946489910949
%x=5, y=0.022977369910025615
The sigmoid function in Equation \ref{eq:RF} has characteristic attributes which are used to quantify RF.
The calculation of the RT can deliver values from zero to $\infty$, which is interpreted as that either the vehicles had an impact, or their paths will never cross in the foreseeable future, respectively.
The sigmoid function for input in $[0; +\infty]$ will deliver values within the interval $]1; 0[$ , which allows us to choose arbitrarily big values for RT when there is no window intersection.
In Figure \ref{fig:riskfinal} we also provide the physical interpretation of the RF.
RF of approx. 0.90 or higher, which correspond to a risk time lower than 1 sec, will be classified as high collision risk.
The value higher than 0.68, which corresponds to RT lower to 2 sec will be classified as high risk.
Further values can be obtained from the figure.

\section{Further Discussion on the Risk Factor}
\label{sec:furtherDiscuss}
When comparing RF with other metrics, it is essential to highlight several key differences. Initially, RF is computed as soon as the Automated Vehicle (AV) detects a VRU. In contrast to the Time To Collision (TTC), RF exhibits a notable distinction in its mathematical representation. It remains definable even in scenarios where there is no potential for collision, that is, when the time intervals for the AV and the VRU do not overlap \((T_{r}^{AV} \cap T_{r}^{VRU} = \emptyset)\).

Furthermore, when contrasting RF with Time To Plan (TTP), another significant difference emerges. TTP assigns a temporal value to all detected traffic participants. Although this feature is valuable, in practical terms, there is no need to account for VRUs or vehicles that will not intersect with the intended path. RF, on the other hand, selectively filters out traffic participants that could potentially pose a collision risk along the planned route.

The methodology outlined in this study is primarily designed for interactions between CAVs and VRUs. However, as previously mentioned, the RF can also be applied to conventional vehicles
The approach involves substituting the anticipated trajectory of the AV with a linear model representing the vehicle's motion. 
Additionally steering freedom can be added, which leads to a coned-shaped prediction of the trajectory.
Subsequently, this predicted trajectory can be superimposed with the legally permissible areas for vehicle navigation. The underlying principle is that vehicles are prohibited from traveling on pedestrian paths or in opposing traffic lanes. Consequently, these sections of the road must be excluded to prevent a high incidence of false positives.

Regarding the application of this metric, there are two primary contexts in which it proves beneficial. Initially, the RF serves as a reliable quantifier of the enhancement in safety attributable to sensing technologies. This implies, for example, that integrating CPS into CAVs equipped with onboard sensors will undoubtedly enhance the RF outcomes. The rationale is that a CAV with connectivity capabilities will detect a VRU as promptly as it would without such connectivity. Consequently, RF is computed only upon the initial detection of a VRU by the CAV. In their standard operation, CAVs are expected to alter their actions immediately when a VRU is perceived within their sensory range and is predicted to be on a collision course, indicating a high RF.
\begin{figure}[!b]
    \centering
    \includegraphics[width=.70\linewidth]{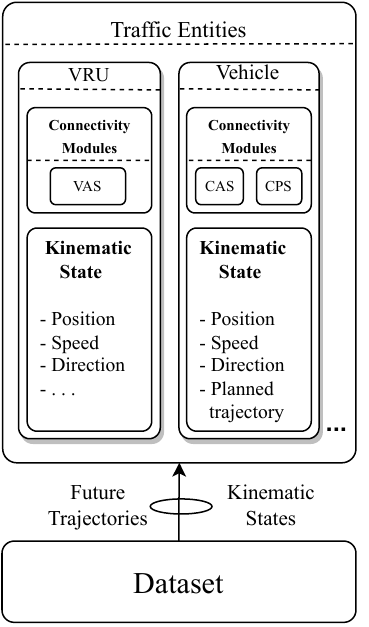}
    \caption{Simulator architecture}
    \label{fig:simArchitect}
\end{figure}
\begin{figure*}[t]
\centering
\begin{minipage}[b]{.48\linewidth}
\includegraphics[width=\linewidth]{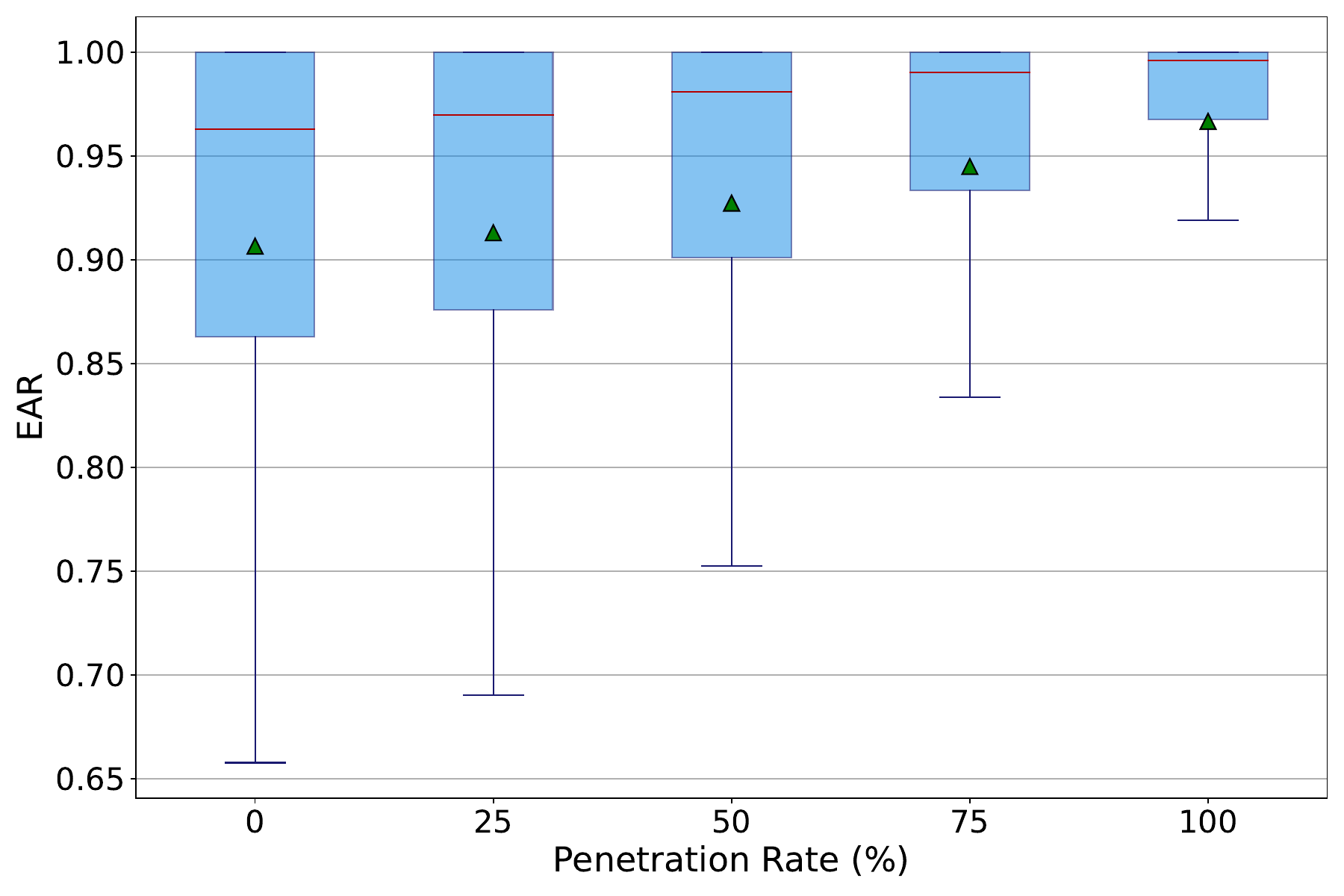}

\caption{EAR for various penetration rates}
\label{fig:ear}
\end{minipage}\qquad
\begin{minipage}[b]{.48\linewidth}
\includegraphics[width=\linewidth]{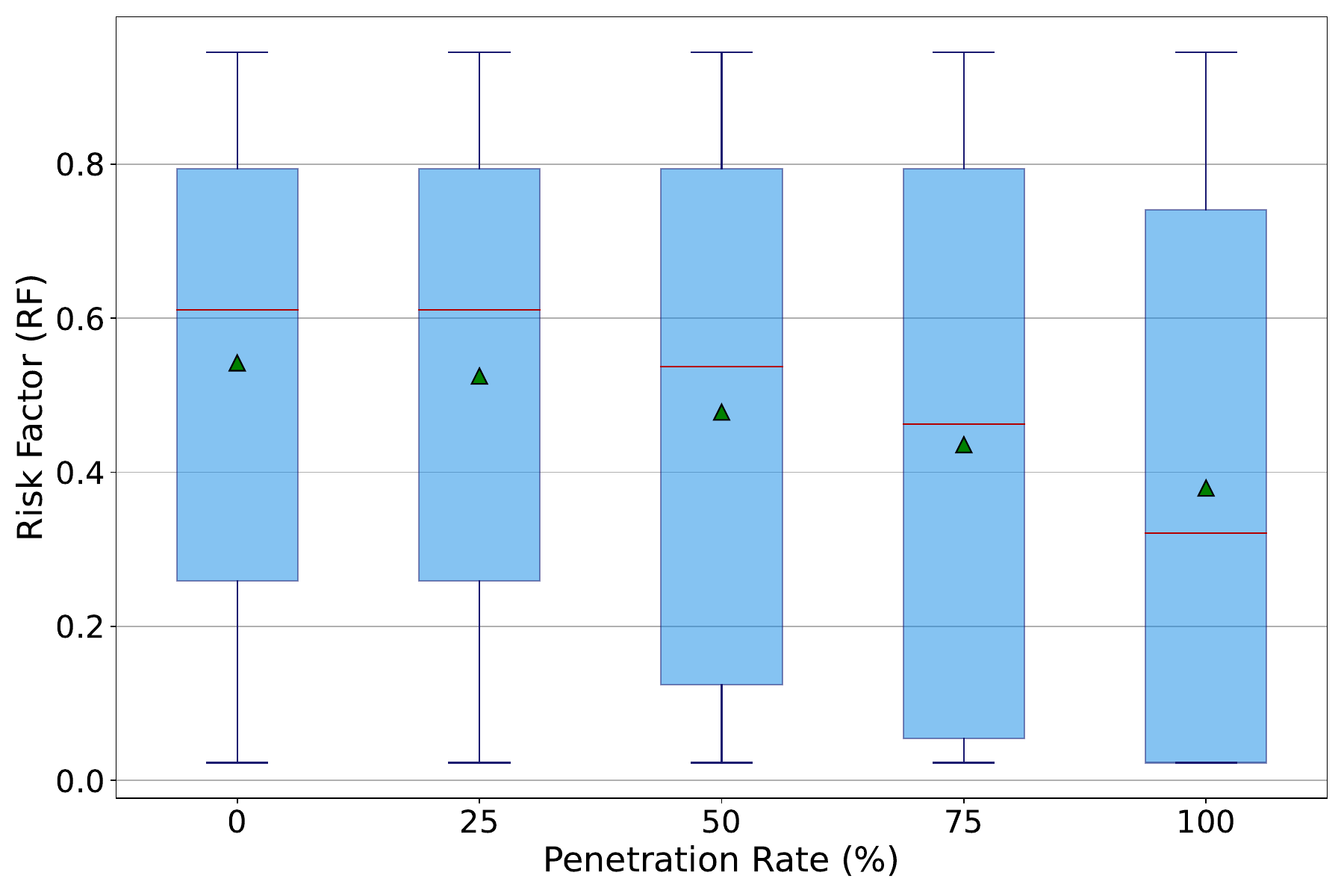}
\caption{Risk Factor for various penetration rates}
\label{fig:rf}
\end{minipage}
\end{figure*}
The second application of this metric pertains to infrastructural analysis. Traffic infrastructure planners can utilize the RF as an indicator of safety during their evaluations. As previously discussed, upon a VRU's entry into a CAV's sensor range, the vehicle should adapt its movement to reduce collision risk. Similarly, human-driven vehicles react to the presence of VRUs or other traffic participants by adjusting driving to mitigate risks. By assessing the RF for legacy vehicles, engineers analyzing specific infrastructures, such as intersections, can identify areas of increased risk, which may arise from obstacles that obstruct the driver's view. The results section will give examples of these applications, showing risk analysis on a real-world dataset and potential risk hotspots depicted from the RF.

\section{Simulation Framework}
\label{sec:simFramewor}

This section introduces the simulation framework used as the evaluation method in this study, depicted in Figure \ref{fig:simArchitect}.
For this study we use the drone-based Intersection drone (InD) dataset \cite{bock2020ind}. It includes recorded data from four different German crossroads. Table \ref{tab:traffic_data} gives an overview of recording data and entities in this dataset. 
The simulator, grounded in real-world data, processes movement and state information to model traffic dynamics accurately.
The traffic participants are categorized into VRUs and vehicles, each with specific kinematic states, potential future paths, and tailored connectivity modules.
Vehicles are enhanced with Cooperative Awareness Service and CPS, incorporating simulated on-board sensors.
These sensors are modeled as 360$\degree$ radial beams enveloping the vehicle, ensuring entities are detected based on beam coverage criteria.
If the beams cover more than half of the surface of another vehicle, or an entire VRU, these are considered as sensed by the ego vehicle.
Upon detection, the identified vehicle or VRU is incorporated into the ego vehicle's local environment model, enabling their subsequent inclusion in the collective perception messages.

\subsection{Simulation Setup}
The simulation uses tracks from the inD dataset as inputs, computing sensor occlusion and connectivity for each frame provided by the inD dataset.
This achieves an update frequency of 25 Hz for all elements in the simulation.
For this scenario, we assume an ideal communication model with no packet loss, where V2X messages are received in the $n^{th} + 1$ frame after they are sent by the originating station.
We will also model different market penetration rates for CAVs in 25\% increments. For low penetration rates, the number of communication entities present in any given frame is expected to be low. This is due to the low traffic density in the dataset: The mean parallel traffic participants present in a frame, calculated over all recordings, is approximately 15. This number includes VRUs and vehicles.

The simulation incorporates the real trajectories of vehicles from the dataset for AV simulation.
Vehicles selected for simulation as CAVs use their future trajectories as the planned AV path.
Our proposed RF metric is evaluated at every frame for each vehicle interacting with VRUs present.
A risk is not considered new if it persists across consecutive frames; according to our RF definition, only the initial non-occluded intersection of risk areas is accounted for in the risk assessment.

\section{Results}
\begin{table}[!b]
\centering
\caption{Summary of traffic data and risk by location}
\label{tab:traffic_data}
\begin{tabular}{|p{0.2cm}|>{\raggedleft\arraybackslash}p{0.9cm}|>{\raggedleft\arraybackslash}p{1cm}|>{\raggedleft\arraybackslash}p{0.8cm}|>{\raggedleft\arraybackslash}p{1cm}|>{\raggedleft\arraybackslash}p{0.8cm}|>{\raggedleft\arraybackslash}p{0.8cm}|}

\hline
\textbf{ID} & \textbf{Duration [min]} & \textbf{Vehicles (Count)} & \textbf{VRUs (Count)} & \textbf{RF Incidences} & \textbf{Mean RF} & \textbf{Stdev} \\ \hline
1           & 185                               & 2503                        & 1235                    & 236                & 0.4115           & 0.2999         \\ \hline
2           & 243                               & 2436                        & 3799                    & 534                 & 0.5901           & 0.2808         \\ \hline
3           & 51                                & 1196                        & 83                      & 13                 & 0.5386           & 0.3553         \\ \hline
4           & 110                               & 2098                        & 249                     & 27                 & 0.7342           & 0.1961         \\ \hline
\end{tabular}
\end{table}

Our simulation results are presented in this chapter.
Different market penetration rates of CAVs are simulated over all recordings and locations provided by the inD dataset.
Table \ref{tab:traffic_data} presents a summary of the dataset we analyzed. It includes tracking data from four different intersections, covering a total of 589 minutes of recorded traffic.
The table reveals that recordings 3 and 4 have fewer data points, including a limited number of VRUs and recorded time.
In contrast, Location 1 and 2 provide a richer dataset, which we will examine more closely in our detailed analysis of these intersections.
However, for our overall assessment of risk metrics, we consider the data from all recordings.

Initially, the Environmental Awareness Ratio (EAR) of vehicles concerning VRUs is assessed.
EAR is calculated as the portion of VRUs known to the ego-vehicle divided by the entire number of VRUs within a radius of 25 meters.
The accompanying Figure \ref{fig:ear} illustrates the EAR distribution for CAVs market penetration rates varying from 0 to 100\%, depicted via box plots.
This percentage reflects the share of CAVs equipped with CPS.
An evident pattern emerges, showing EAR enhancement as the penetration rate ascends. Given the focus on VRUs proximal to vehicles, elevated EAR values are anticipated due to the comprehensive coverage of VRUs by the CAVs' onboard sensors.
Consequently, a scenario with no CPS-equipped vehicles yields a median EAR of 0.96, which increases to 0.99 at a 100\% penetration rate.
More pronounced trends are observed in the lower quartile and whisker values, increasing from 0.865 to 0.97 (lower quartile) and 0.66 to 0.92 (lower whisker), respectively.
It is noteworthy that the EAR, also referenced as the perception ratio in analogous studies, does not exhibit as marked an increase as identified by other researchers.

In Figure \ref{fig:rf}, the RF is illustrated for various CAV market penetration rates.
Across all rates, RF values exhibit a broad range, spanning from values slightly above 0 to below 1.
Similar to the evaluation of EAR, an evident trend of decreasing RF is observed as the market penetration rate increases; however, unlike the EAR distribution, the spectrum of RF values remains wide.
This results in a drop in the median RF from 0.61 (at 0\% penetration rate) to 0.34 (at 100\% penetration rate), a relative decrease of 44\%. It is noteworthy that a penetration rate of 25\% does not yield any RF improvement. This phenomenon can be attributed to the marginal rise in EAR, as depicted in Figure \ref{fig:ear}. Given the traffic density in the dataset, there is a high likelihood of having no or just one CAV present at any time. However, at a 100\% penetration rate, a notable reduction in risk is observed in the upper quartile, with the 75th percentile decreasing to 0.76 and the average risk falling to around 0.4.
These values are equivalent to the risk time of approximately 2 and 3 seconds respectively.

\begin{figure}[t]
  \centering
  % First subfigure
  \begin{subfigure}[b]{\linewidth}
    \includegraphics[width=\linewidth, trim=0cm 1.4cm 0cm 2.0cm, clip]{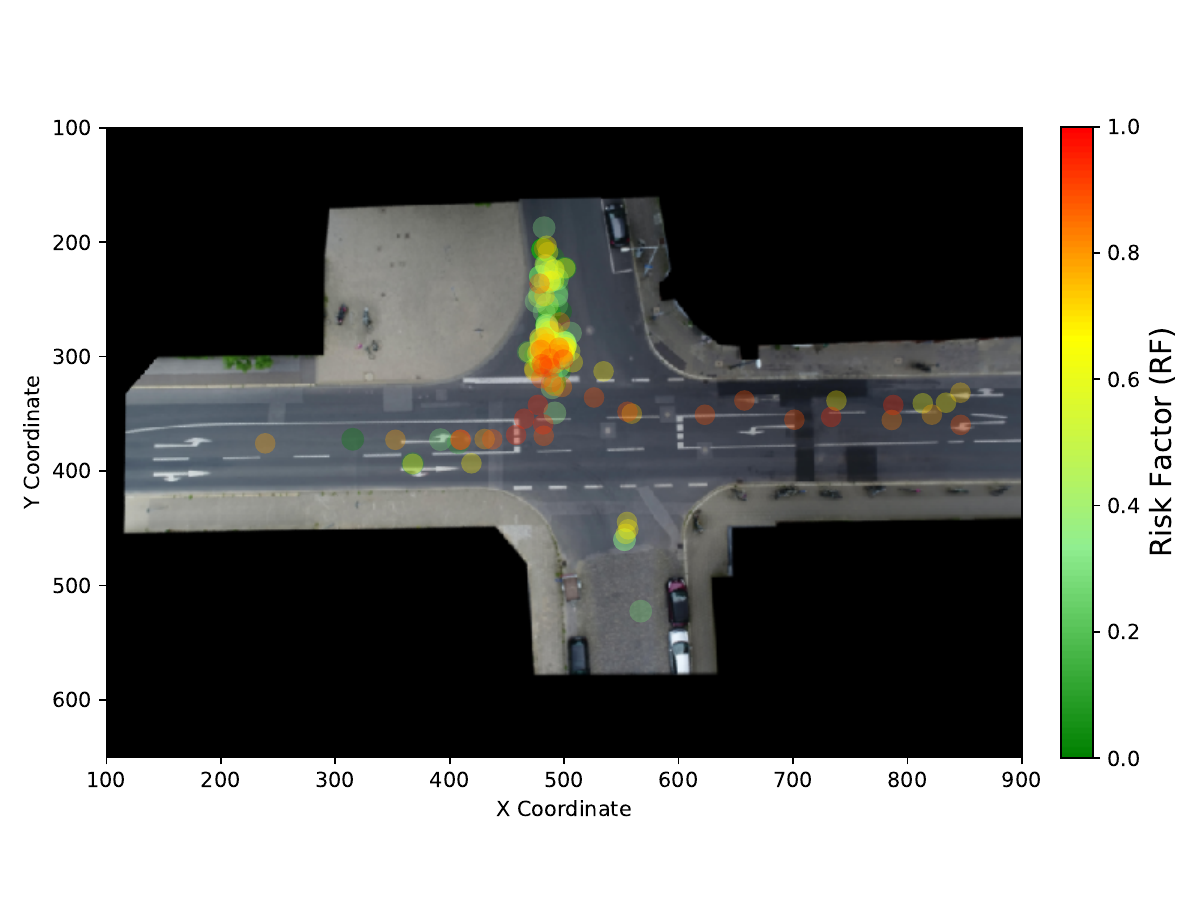}
    \caption{Location 1 Crossroad}
    \label{fig:heatmap1}
  \end{subfigure}
  \begin{subfigure}[b]{\linewidth}
    \includegraphics[width=\linewidth, trim=0cm 1.4cm 0cm 2.0cm, clip]{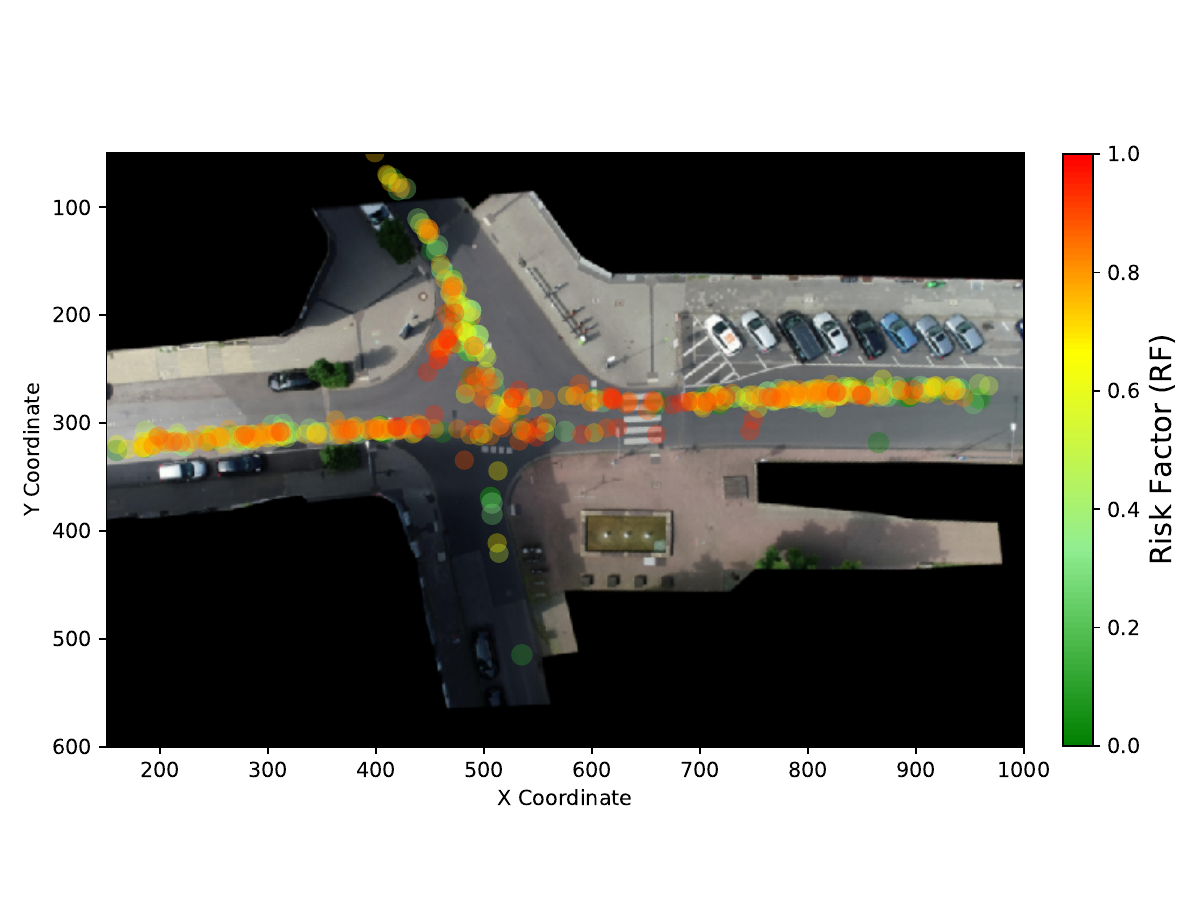}
    \caption{Location 2 Crossroad}
    
    \label{fig:heatmap2}
  \end{subfigure}
  \caption{Risk heat map for two different crossroad}
  \vspace{-0.4cm}
\end{figure}

After reviewing the risk assessment for varying levels of CAV market penetration, we focus in on two specific areas, locations 1 and 2, which provide the most significant amount of data as shown in table \ref{tab:traffic_data}. 
This analysis offers detailed insights into the RF metric, mapping out the locations where risks are most likely to occur. It highlights potentially hazardous spots at intersections, offering a clearer understanding of where and how dangers emerge.
To analyze this, we create a heatmap showing the RF incidents in these areas. The heatmap uses a color scale from green to red to show increasing risk, with the RF value representing the level of risk. The placement of the dots on the heatmap shows the location of the vehicles at the moment it encountered a potential risk with a VRU. In location 1 (Figure \ref{fig:heatmap1}), there's a noticeable cluster of high-risk incidents at the top arm of the intersection. This is the intersection arm where most vehicles enter the main street, with their planned trajectories directed towards either the left or right intersection arm. This area is also where most vehicles and VRUs travel, thereby increasing the risk of VRU occlusion and intersection of VRU trajectories.

Location 2 is depicted in Figure \ref{fig:heatmap2}, where there's a significant concentration of risk on the left, top, and right sides of the intersection. It's evident that the risk increases as one gets closer to the center of the intersection. Overall, the risk is more uniformly spread throughout the intersection in location 2. Additionally, parked cars on the right side of the intersection block the view of VRUs, leading to an average risk level of 0.59, as detailed in table \ref{tab:traffic_data}.
It is noteworthy that in none of the locations RF reaches 1.0 values, as that would mean an impact between vehicles and VRUs.

\label{sec:results}

%\begin{figure*}
%\centering
%\begin{minipage}[b]{.4\textwidth}
%\includegraphics[width=\linewidth]{figures/ear.pdf}
%    \label{fig:ear_high}
%\caption{Caption}\label{label-a}
%\end{minipage}\qquad
%\begin{minipage}[b]{.4\textwidth}
%\includegraphics[width=\linewidth]{figures/risk_violin.pdf}
%    \label{fig:ear_high}
%\caption{Caption}\label{label-b}
%\end{minipage}
%\end{figure*}

\section{Conclusions}
\label{sec:conclusions}
The proposed Risk Factor (RF) enhances our understanding of safety in real-world scenarios by identifying critical risk hotspots through the analysis of actual vehicle and VRU trajectories.
Risk hotspots can be identified and road planners may adapt properly, for instance by placing traffic lights.
Moreover, the RF highlights the advantages of V2X communication in mitigating risk to VRUs.
Unlike other metrics, such as the EAR, which primarily focuses on measuring the perception ratio for Connected Vehicles (CVs), the RF showcases the effect of communication on the potential overlap of trajectories in the forthcoming seconds.
In real-world traffic scenarios, as provided by inD dataset, the impact of V2X on risk, quantified by RF, is lower than the improvement of EAR, indicating the need for safety-centered evaluation metrics.
These insights help to improve VRU safety estimation, offering a more comprehensive view of how existing infrastructure and communication can be designed for safety improvement.

\section*{Acknowledgment} This publication was funded by the Deutsche Forschungsgemeinschaft (DFG, German Research Foundation) - project number 227198829 / GRK1931 and by the Lower Saxony Ministry of Science and Culture under grant number ZN3493 within the Lower Saxony “Vorab“ of the Volkswagen Foundation and supported by the Center for Digital Innovations (ZDIN).

\bibliographystyle{ieeetr}       
\bibliography{references}

\end{document}